\documentclass[aps,prb,twocolumn,groupedaddress,superscriptaddress,showpacs,amssymb,amsmath]{revtex4-2}
\usepackage{graphicx}
\usepackage{amsmath}
\usepackage{mathtools}
\usepackage{amssymb}
\usepackage{amsfonts}
\usepackage{color}
\usepackage{bm}        
\usepackage{comment}
\usepackage{placeins}
\usepackage[mathscr]{eucal}
\usepackage[colorlinks, linkcolor=red,citecolor=blue,urlcolor=blue]{hyperref}
\usepackage[normalem]{ulem}
\usepackage[all]{hypcap}
\usepackage{multirow}
\usepackage[dvipsnames,usenames,table]{xcolor}
\usepackage{dcolumn}
\usepackage{hyperref}
\usepackage{bm}
\usepackage{epsf}
\usepackage{subfigure}
\usepackage{amsfonts}
\usepackage{epstopdf}%
\usepackage{braket}
\usepackage{float}

\newcommand{\be}{\begin{equation}}
\newcommand{\ee}{\end{equation}}
\newcommand{\bea}{\begin{eqnarray}}
\newcommand{\eea}{\end{eqnarray}}
\begin{document}

\title{Cusp singularities in the  distribution of orientations of  asymmetrically-pivoted  hard discs on a lattice}
\author{Sushant Saryal}
\email{ss0410@princeton.edu (corresponding author)}
\affiliation{Department of Physics, Indian Institute of Science Education and Research, Pune 411008, India}
\affiliation{Department of Chemistry, Princeton University, Princeton, New Jersey 08544, USA}
\author{Deepak Dhar}
\email{deepak@iiserpune.ac.in}
\affiliation{Department of Physics, Indian Institute of Science Education and Research, Pune 411008, India}
		
\date{\today}

\begin{abstract}
We study a system of equal-sized circular discs each with an asymmetrically placed pivot at a fixed distance from the center. The pivots are fixed at the vertices of a regular triangular lattice. The discs can rotate freely about the pivots, with the constraint that no discs can overlap with each other.  Our Monte Carlo simulations show that the one-point probability distribution of orientations have multiple cusp-like singularities. We determine the exact positions and qualitative behavior of these singularities.  In addition to these geometrical singularities, we also find that the system shows order-disorder transitions, with a disordered phase at large lattice spacings, a phase with spontaneously broken orientational lattice symmetry at small lattice spacings, and an intervening Berezinskii-Kosterlitz-Thouless phase in between.
\end{abstract}

\maketitle 

\section{Introduction}

In many molecular solids, the melting transition from the low-temperature crystalline solid phase to the high-temperature liquid phase does not occur in a single step, but one finds a multiplicity of mesophases.
These are called `liquid crystals', or `plastic solids'. In the former, the periodic three-dimensional crystal structure is absent, but varying amount of orientational order may be present. In the latter, the average positions of the centers of masses of molecules do lie on a three-dimensional crystalline lattice, but there is none or only partial orientational order.  These were originally called plastic solids, as they can be deformed easily using much less force, compared to `hard' crystals. Some examples of common materials that show plastic solid phases are nitrogen\cite{nitrogen}, carbon tetrabromide \cite{powell}, formylferrocene \cite{kaneko}. The currently favored nomenclature for these is orientationally disordered crystals. In recent years, these have attracted a  lot of interest, because of their promising applications in diverse areas such as solid electrolytes\cite{batteries}, drug delivery\cite{drug-delivery}, optoelectronics\cite{opto}, barocalorics\cite{refrigeration}, piezoelectrics\cite{piez}, etc. For a recent review of the applications, see Das {\it{el al}}\cite{Das}.

Pauling in 1930  derived a rough criteria for strongly hindered rotational motion of molecles in crystalline solids \cite{pauling}. But it was Timmermans who systematized the phenomenological study of plastic crystals starting from the 1930s\cite {timmermans}. 
On the theoretical front, Pople and Karasz \cite{pople-karasz} extended the two-lattice model of Lennard-Jones and Devonshire\cite{Lennard} to account for the order-disorder transition in the orientation of molecular crystals. A  minimal model for these would be  to assume the constituents as rigid objects each identically pivoted on a lattice and free to rotate provided no objects overlap with each other.  Casey and Runnels\cite{runnels1} and Freasiers and Runnels \cite{runnels2} examined a system of hard squares with centers ﬁxed on the 1d lattice. 
We have recently discussed this model, and called it rigid hard rotors on a lattice as  model of multiple phases shown by plastic crystals to describe the transitions between them \cite{dd1, dd2,dd3}. Note that  since the lattice is always present, the model does not have a `liquid' phase with no crystalline order. 

In an earlier paper \cite{dd3},  we determined the exact functional form of the one-point probability distribution function of orientations at a site for a range of lattice spacings when a particular condition, called the at-most one overlap (AOO) condition, holds. In this paper we particularly examine a system of hard discs asymmetrically pivoted on a triangular lattice and study the one-point probability distribution function (PDF) $P(\theta)$ of orientations $\theta$  beyond the AOO condition. We find that the distribution  function $P(\theta)$ shows cusp singularities. We determine  the position and qualitative behaviour beyond the AOO condition exactly. Singularities in the pair distribution function were studied earlier by Stillinger\cite{still1} and numerically observed in the the probability distribution of bond-pair angles in a system of hard spheres \cite{still2}, but non-trivial  singularities in the one-point function  have not been discussed before. We also numerically verify our findings with the help of Monte Carlo simulations.\\  

This paper is organized as follows. In section \ref{model} we define our model. In section \ref{distribution} we show that there exist multiple cusp singularities in the one-point probability distribution of orientations and exactly determine their nature and  positions. In section \ref{MCR} we verify our findings using Monte Carlo simulations. Section \ref{remarks}  contains some concluding remarks.

\section{Model}\label{model}

\begin{figure}[htp]
\centering
\includegraphics[width=0.8\columnwidth]{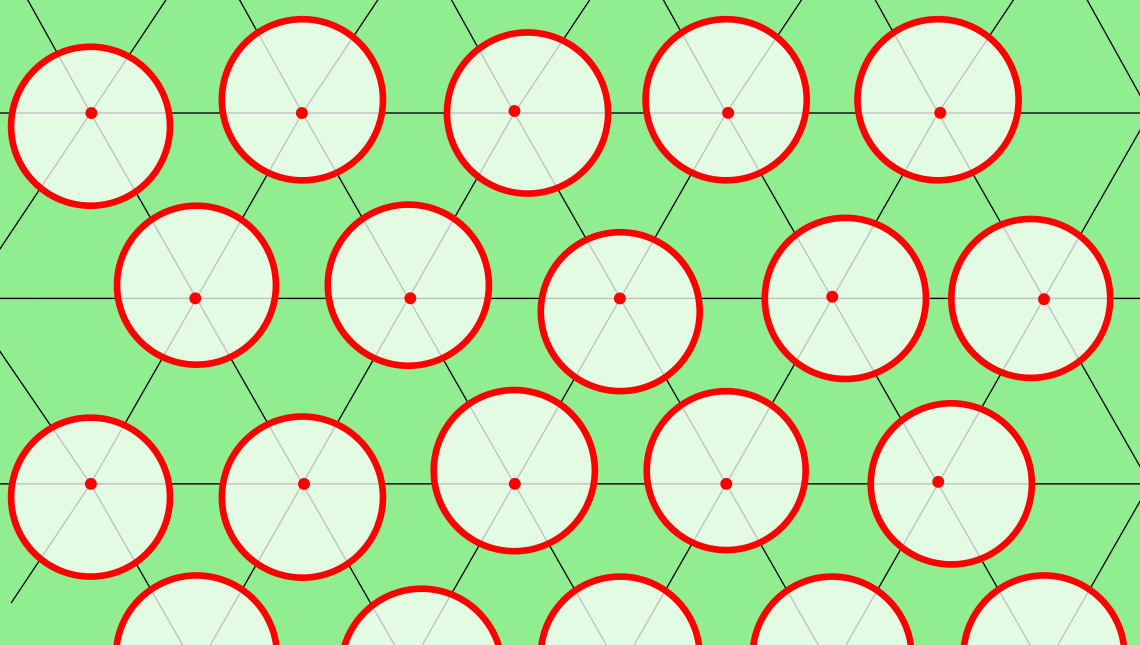}
\caption{A set of unit-radius hard circular discs pivoted asymmetrically on the triangular lattice with lattice spacing $a$. Pivot is placed at a distance $\epsilon$ from the center.}
\label{fig1}
\end{figure}

We consider a system of identical unit-radius circular discs each with an asymmetrically placed pivot at a distance $\epsilon$ from the center, as shown in Fig \ref{fig1}. The pivots form a regular triangular lattice with lattice spacing $a$, and the discs can rotate freely about the pivots, with the constraint that no discs can overlap with each other. Orientation of a disc pivoted at lattice site $\bf{r}$ is specified by an angle $\theta(\bf{r})$, measured between long-axis (passing through the pivot and center) and $x$-axis. From elementary geometry it is clear that if $a > 2(1 + \epsilon)$, the discs rotate freely, and the closest-packing limit is reached when $a = 2$. We write lattice spacing $a$ as, 
\bea
a=2+\epsilon x,
\label{lattce}
\eea
with $0 \leq x \leq 2$.\\
The partition function of the system  having $N$ discs is  
\bea
\mathcal{Z}_{N} = \left[ \prod_{\bf{r}} \int_{-\pi}^{\pi} \frac{d\theta(\bf{r})}{2\pi} \right]  \prod_{\mathclap{\substack{\mathbf{x},\mathbf{y} \\ \mathbf{x}\ne\mathbf{y}}}} \left[ 1 - \eta\big(\theta(\bf{x}),\theta(\bf{y})\big)\right],
\label{eq:Z}
\eea
where $\eta\big(\theta(\mathbf{x}),\theta(\mathbf{y})\big)$ is an indicator function which is one when discs at sites $\mathbf{x}$ and $\mathbf{y}$ with orientations $\theta(\bf{x})$ and $\theta(\bf{y})$ respectively, overlap and zero otherwise.  We define the entropy per site $s(x,\epsilon)$ by 

\bea
s(x,\epsilon) = \lim_{N \rightarrow \infty} \frac{ \log \mathcal{Z}_N} { N}
\eea

\begin{figure}[htp!]
 \includegraphics[width=0.35\textwidth]{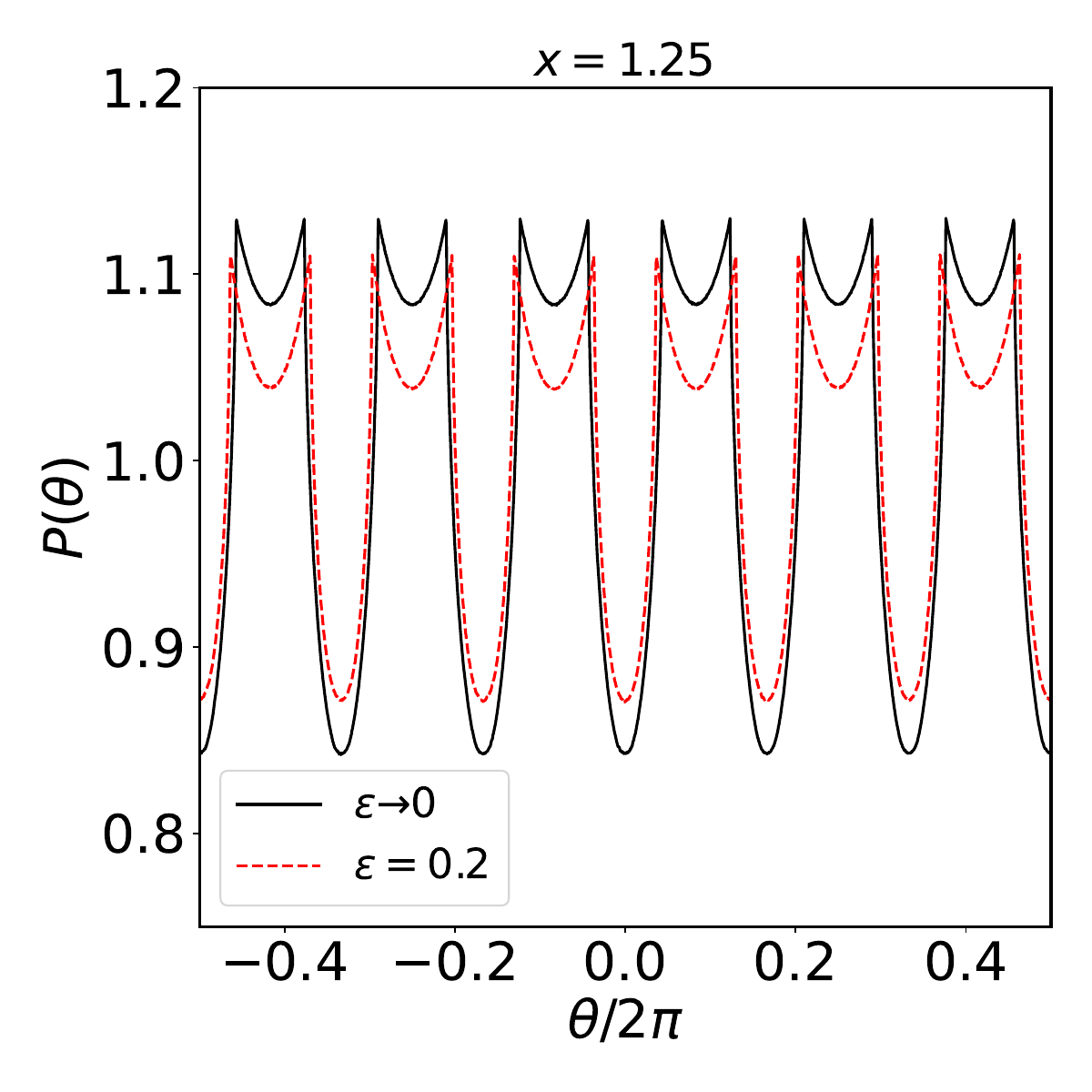}
    \caption{Comparison of $P(\theta)$ for finite $\epsilon$ and $\epsilon \to 0$ at $x=1.25$.}
    \label{finite}
\end{figure}

As $\epsilon$ tends to zero, the function $s(x,\epsilon)$ has a well-defined non-trivial limit
\bea
s(x) = \lim_{\epsilon \rightarrow 0} s(x,\epsilon).
\eea
In this limit, the no-overlap condition between two neighboring  sites simplifies. 
For two neighboring rotors, if the line  joining the pivots  makes an angle $\phi$ with the $x$-axis, the no-overlap condition to first order in $\epsilon$ becomes
\bea
x-\cos (\theta-\phi) + \cos (\theta^{'}-\phi) \ge 0.
\label{ovlap2}
\eea
In Fig. \ref{finite}, we compare the plots of $P(\theta)$ for $\epsilon \to 0$ and $\epsilon =0.2$ using Monte-Carlo simulations. We see that the qualitative behaviour of $P(\theta)$ is same for the two cases.

The limit $\epsilon \rightarrow 0$ has the advantage that the number of parameters specifying the model is reduced to $1$. In the following, for the sake of simplicity, we restrict our discussion to this case. The case of more general $\epsilon$ presents no additional special features as evident from Fig. \ref{finite}.

This model can also be thought of as a system of planar spins $\{\theta(\bf{r})\}$ on the vertices of a triangular lattice, with nearest neighbor interaction Hamiltonian $\mathcal{H}$ given by,
\bea
\mathcal{H}=J\sum_{\bf{r}} \sum_{j=0}^{2}\Theta \left[\cos( \theta(\mathbf{r}) - j \pi/3)-\cos(\theta_j^{'}(\mathbf{r})-j\pi/3) -x\right] \nonumber \\ 
\eea
where $\theta_j^{'}(\bf{r})$ is  the  neighboring spins of the spin $\theta(\bf{r})$ in the lattice direction  $j \pi /3$, and  $\Theta(x)$ is the Heavyside step function of $x$.  The hard-core limit corresponds to setting $J$ to $+ \infty$.

This model is of the same form as the model of hard-core spins studied earlier by Sommers et al \cite{sommers}. These authors studied the case where our condition Eq.(\ref{ovlap2}) is replaced by $|\theta -\theta'| \leq \alpha$. The qualitative behavior of the models is similar. The main difference where our model differs from theirs is the explicit breaking of isotropy in the spin space by the lattice-direction dependent interaction. In particular, the cusp singularities we discuss here are not present in the hard-core spin model.  

\begin{figure}[htp!]
    \centering
    \includegraphics[width=0.42\textwidth]{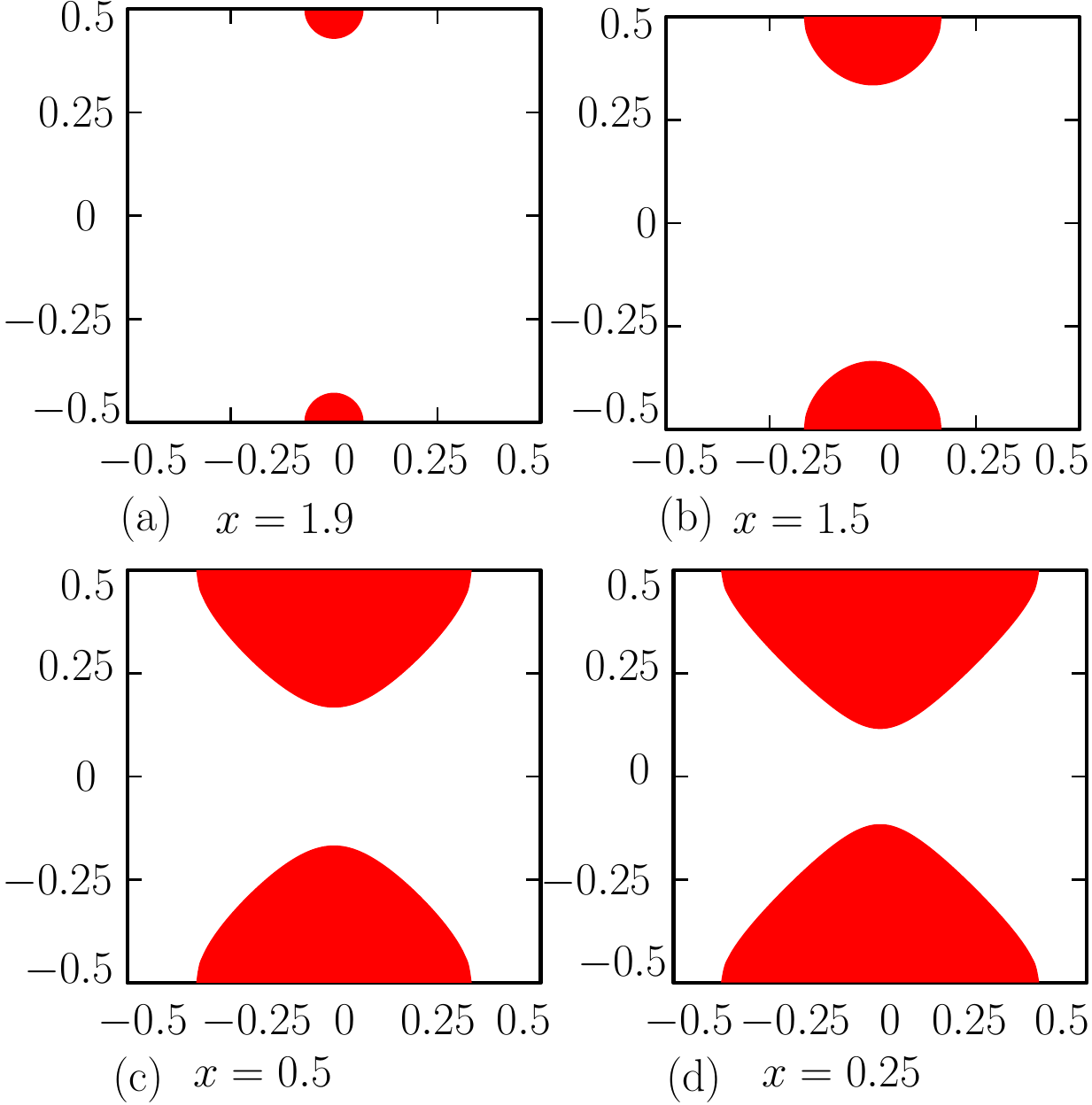}
    \caption{$\eta(\theta,\theta^{'})$ for nearest-neighbor discs along x-axis for different values of $x$ when $\epsilon \to 0$}  in \Big($\dfrac{\theta}{2 \pi}$,$\dfrac{\theta^{'}}{2 \pi}$\Big) plane. Shaded area (red) corresponds to overlap region where $\eta$ is  $1$. In the unshaded area $\eta =0$. For other nearest-neighbor pairs, $\eta(\theta,\theta^{'})$ can be similarly obtained using the symmetries of the triangular lattice.
    \label{tm1}
\end{figure}

\section{Probability distribution of orientations}\label{distribution}
Let $P(\theta) d\theta$ denote the probability  that the disc pivoted to a randomly picked  site in the  equilibrium ensemble described by the partition function in Eq. (\ref{eq:Z}) is at an orientation between $\theta$ and $\theta +d\theta$.

\subsection{At-most One Overlap(AOO)}
The simplest case is when $x$ lies in the range $[1+\sqrt{3}/2,2]$. In this case, it is easily seen that in any configuration of orientations of discs, any disc can overlap with at-most one other disc. It was called the AOO condition in our earlier paper\cite{dd3}. We showed that, within AOO regime, the partition function simplifies to the calculation of the partition function of dimers and vacancies on the same lattice with the activity of dimers is given by
\bea
z=-\int_{-\pi}^{\pi} \frac{d\theta}{2\pi} \int_{-\pi}^{\pi} \frac{d\theta^{'}}{2\pi} \eta(\theta,\theta^{'}).
\eea
We also obtained an explicit expression for $P(\theta)$, involving an undetermined function $\bar{n}(z)$, which gives the number density of dimers $\bar{n}(z)$, as a function of their activity $z$ within AOO regime in $d$ dimensions. Following this general expression, the one-point PDF in the present scenario is given by 
\bea
P(\theta)= (1-2\bar{n}(z))[1 + \frac{\bar{n}(z)}{3z(1-2\bar{n}(z))} \sum_{i=1}^{6} f_{i}(\theta)]
\label{eq:Ptheta}
\eea
where $f_i(\theta)$ is given by 

\bea
f_i(\theta)=\int_{-\pi}^{\pi} \frac{d\theta_{i}}{2\pi} \eta(\theta,\theta_{i})
\eea
and $\bar{n}(z)$ is the dimer number density at activity $z$, which is given by the low density expansion

\bea 
\bar{n}(z) = \sum_{n=1}^{\infty} (-1)^{n-1} a_n  z^n,
\eea
where $a_n$ is the number of heaps made of dimers \cite{vien}.
For the triangular lattice, we have
\bea
\bar{n}(z) = 3[z-11z+144z^2-2047z^3+30526z^4....\,]
\eea
In our problem, the explicit expression for the function $\eta(\theta,\theta_{1}) $ is 
\begin{widetext}
\bea
\eta(\theta,\theta_{1})=
\begin{cases}
        1, & \text{if } |\theta| \le \arccos(x-1) \,\,\,\text{and}\,\,\, |\theta_1| \ge \arccos[\cos(\theta)-x].  \\
        0, & \text{otherwise}.
    \end{cases}
\eea 
\end{widetext}

Then, it is easily seen that  $f_1(\theta)$ is given by 
\bea
f_1(\theta)=
\begin{cases}
         2(\pi-\arccos[\cos(\theta)-x]), \,\, |\theta| \le \arccos(x-1).\\
         0, \,\,\,\,\,\,\,\,\,\,\,  \text{otherwise}.
\end{cases}
\label{ft1}
\eea
Other $f_i$'s can be easily found using the symmetries of the underlying lattice. From this, it is easily seen that $P(\theta)$ has square-root cusp singularities (see Appendix A for more details) at  

\bea
\theta_{\text{cusp}}=  \frac{j\pi}{3}\, \pm \,\arccos(x-1)  
\label{eq:cusppositions}
\eea
where $j=0, 1, 2, .... 5$. 

\begin{figure}[htp]
    \includegraphics[width=0.45\textwidth]{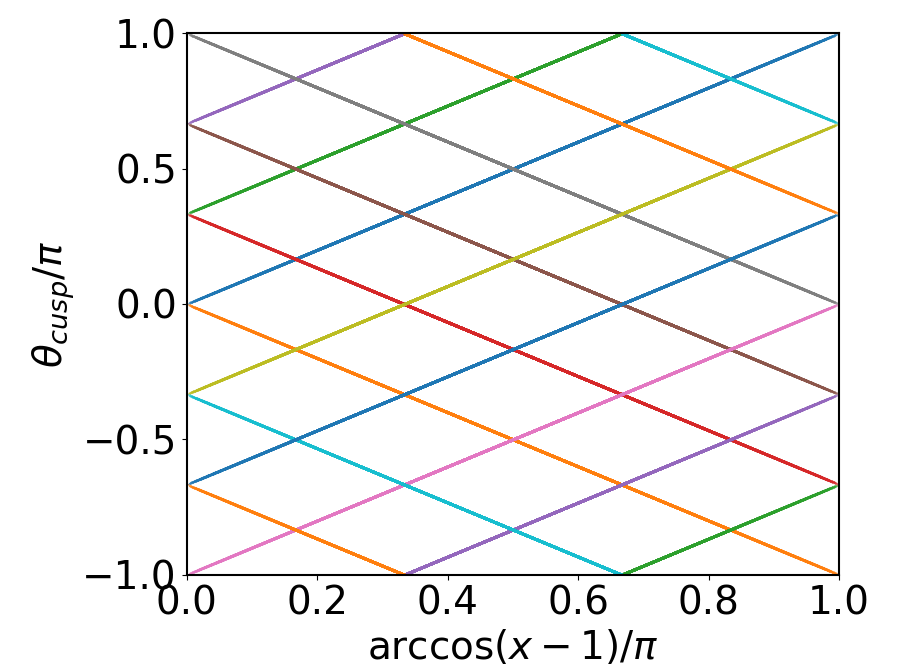}
    \caption{The change in the position of cusp singularities in $P(\theta)$ as  we vary $x$.  The positions vary linearly with the variable  $\arccos(x-1).$ }  
    \label{cuspfig}
\end{figure}

\subsection{Beyond the AOO condition}
Now, we consider $x$ outside the regime where the AOO condition holds. When $x = 1 + \sqrt{3}/2 - \delta$, with $\delta $ positive, but small, the AOO condition is no longer satisfied. In this case, one can still define the graphical expansion of Eq.(\ref{eq:Z}) in terms of configurations of dimers, but now, the configurations where two or more dimers are incident on a vertex have non-zero weight. However, if $\delta$ is small, the weights of such  vertices is small. This suggests that we can organize the terms of this series in the following form

\bea
\mathcal{Z} = \mathcal{Z}_0 + \mathcal{Z}_1 + \mathcal{Z}_2 +...
\eea

In this expansion, $\mathcal{Z}_r$ is the sum over terms having $r$ dimer pairs  such that the dimers in each pair have a common vertex.  We may associate an extra  factor $y$ with each such overlapping dimer pair, and consider the partition sum
\bea
\mathcal{Z}(y) = \sum_{r=0}^{\infty} y^r \mathcal{Z}_r.
\eea
We  treat $y$ as a small parameter, and assuming the sum converges for small enough $y$, treat it is  as 
a perturbation series in $y$.  If we put $y =0$ in this series, we have a sum over the all dimer configurations. In this ensemble, one can define the one-point function $P(\theta)$ as 

\bea
P(\theta) =  \sum_{\bf{x}}\langle\delta(\theta(\bf{x}) -\theta) \rangle
\eea
where the angular brackets denote the average over the  ensemble. It is easy to see that in the unperturbed ensemble $\mathcal{Z}_0$, Eq. (\ref{eq:Ptheta}) continues to remain valid, but now in this ensemble there are ranges of $\theta$ where more than one of the $f$-terms contributes in the equation. And the positions of the cusp singularities in $P(\theta)$ are still given by Eq.(\ref{eq:cusppositions}). 

Consider that the terms in  $\mathcal{Z}_1$,  involving two specified dimers meeting at a specific site $\bf{x}$. Say the dimers are covering the bonds $(\bf{x}, \bf{y})$ and $(\bf{x},\bf{z})$. This weight can be written as a product of two terms $T_1$ and $T_2$, with 
\bea 
T_1 = \int d\theta(\mathbf{x}) \int d\theta(\mathbf{y}) \int d\theta(\mathbf{z}) ~~~\eta(\bf{x},\bf{y}) \eta( \bf{x,}\bf{z})
\eea

It can be shown that for small positive $\delta$, $T_1$ varies as\cite{dd3} $\delta^{3/2}$. $T_2$ is a polynomial in $z$, the sum over all possible partial dimer coverings of the lattice, not involving sites $\bf{x},\bf{y},\bf{z}$. Similar statement is valid for higher $r$.  

If we expand the function $P(\theta)$ in powers of $y$, each term in the perturbation  sum is well-behaved, and only singularities in $\theta$ come from integration of the functions $\eta$, and hence the positions  are same as in Eq.(\ref{eq:cusppositions}) and are robust.  By analytic continuation on $y$, we expect these results to hold for all $y$, and hence at $y=1$. Thus, we conjecture that the cusp singularities $P(\theta)$ are given by Eq. (\ref{eq:cusppositions}) for all $x$, in the range $0 < x < 2$. This is shown in Fig. \ref{cuspfig}. Thus, the positions of singularities do not change so long as we work within any finite order of the perturbation theory in $y$. Numerical evidence of this conjecture based on Monte-Carlo simulation is presented next.
  
\begin{figure}[htp]
    \centering
    \includegraphics[width=0.38\textwidth]{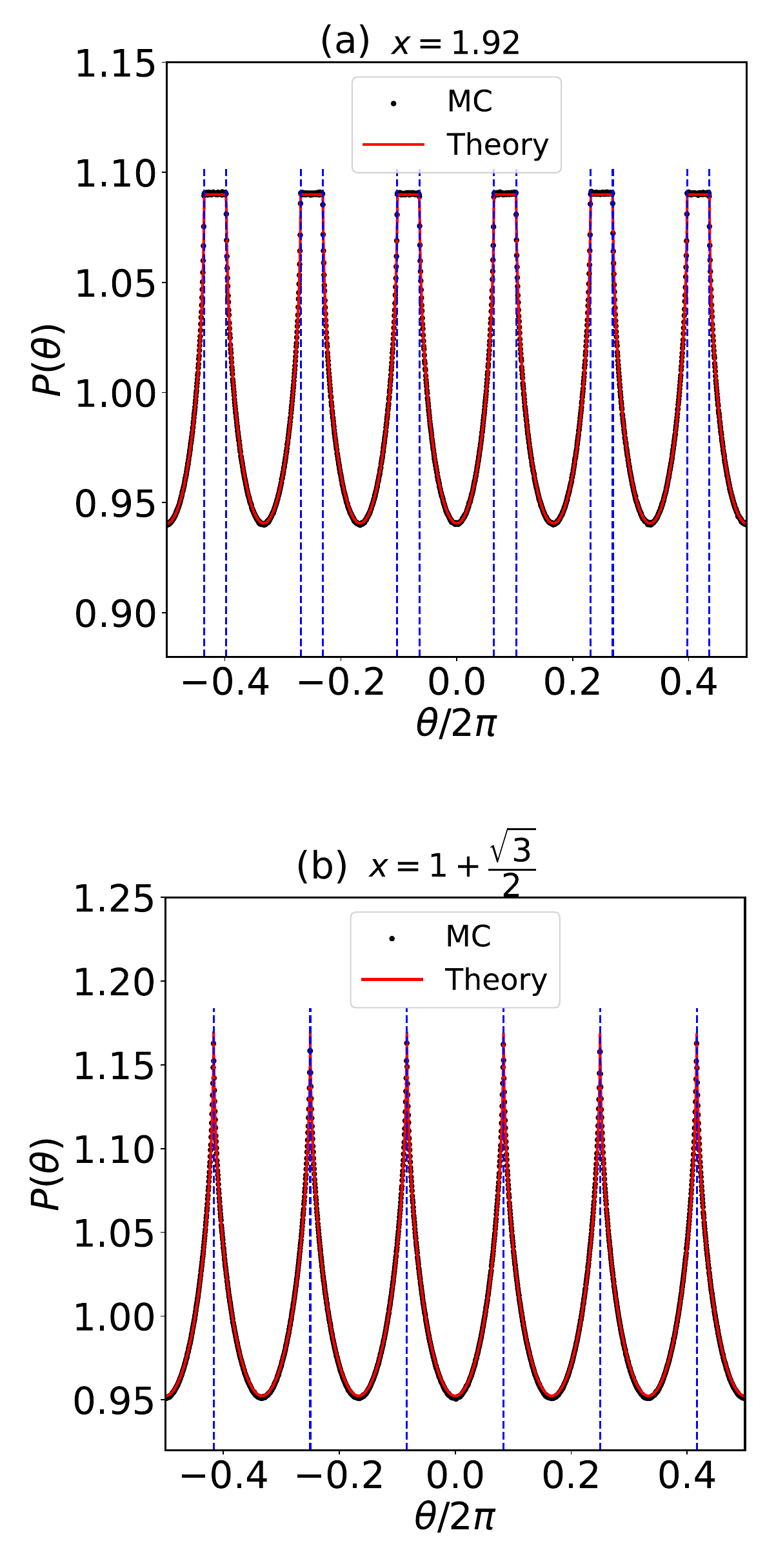}
    \caption{One-point PDF under AOO regime. (a) $x=1.92$ and (b) $x=1+\frac{\sqrt{3}}{2}$. Dashed vertical lines represent the positions of cusp singularities given in Eq.(\ref{eq:cusppositions}). Theoretical predictions, Eq.(\ref{eq:Ptheta}) and Eq.(\ref{eq:cusppositions}), and Monte Carlo results are in very good agreement.}
    \label{aoo1}
\end{figure}  

\begin{figure*}[t]
    \centering
    \includegraphics[scale=0.35]{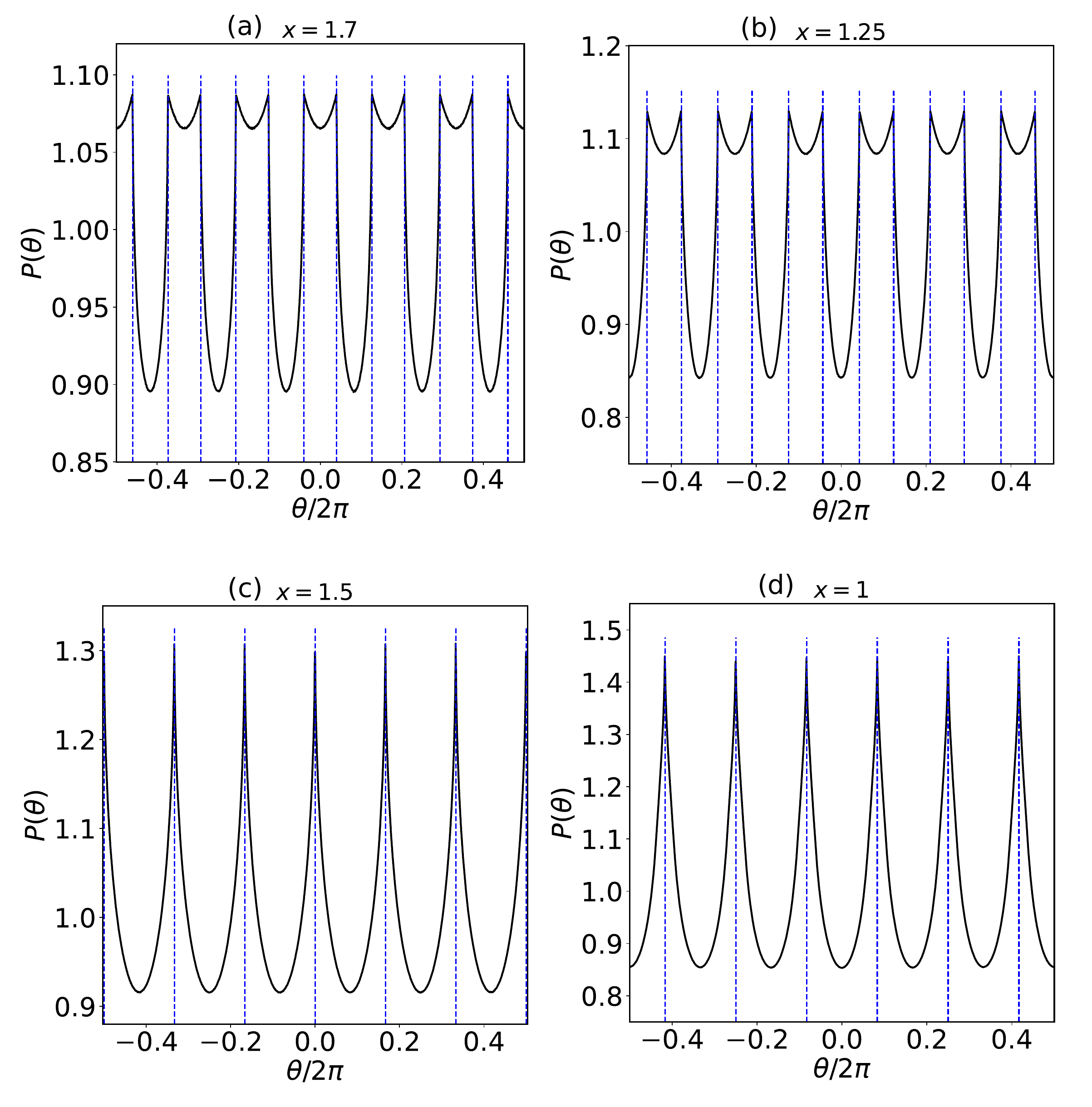}
    \caption{One-point PDF beyond AOO regime. (a) $x=1.7$, (b) $x=1.5$, (c) $x=1.25$, and (d) $x=1$. Dashed vertical represent the positions of cusp singularities given in Eq.(\ref{eq:cusppositions}). }
    \label{baoo1}
\end{figure*}

\section{Monte Carlo Results}\label{MCR}
Now we present our findings of Monte Carlo simulations. Our simulations were done on lattices of size varying from $50 \times 50$ to $100 \times 100$.  We use a single site update scheme: we pick a site $\bf{r}$  at random, and try to change the value of $\theta_{\bf{r}}$ to $\theta_{\bf{r}} + \Delta \theta$, where $\Delta \theta$ is a random variable with a uniform distribution from $-\Delta_0 $ to $+\Delta_0$. We accept the move if the new value does not result in any overlap. And repeat. We average over times of order $7.2$ million MCS, after rejecting the first $8\times 10^5$ steps.

\begin{figure}[htp]
    \centering
    \includegraphics[width=0.38\textwidth]{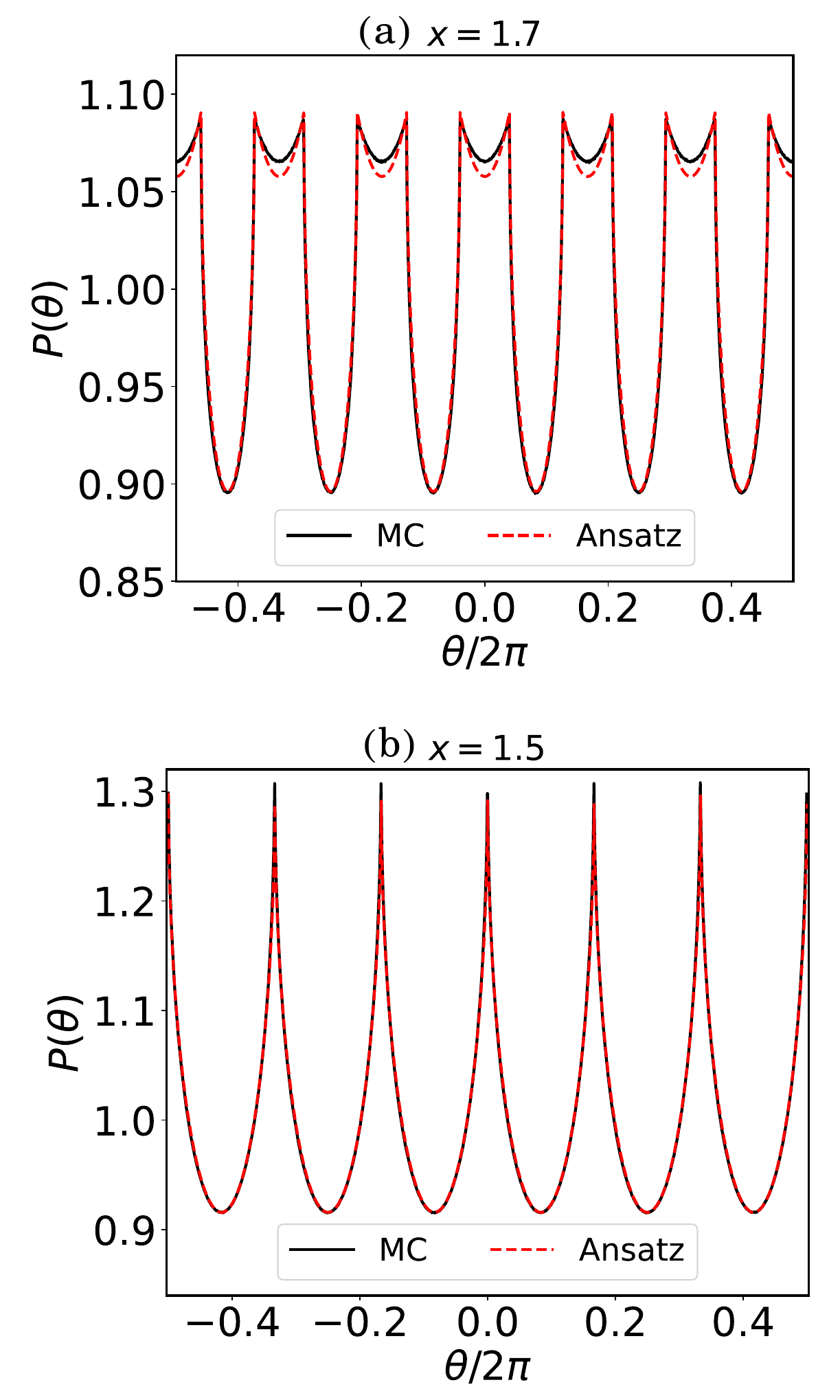}
    \caption{Comparison of one-point PDF obtained using Monte Carlo simulations and our ansatz given in Eq. (\ref{ansatz}).  (a) $x=1.7$, $m=1.405$ and (b) $x=1.5$, $m=1.32$.}
    \label{baoo2}
\end{figure}

In Fig. \ref{aoo1} one-point PDF is plotted using Monte Carlo simulations along with our theoretical prediction, Eq. (\ref{eq:Ptheta}), in the  AOO regime  $ 1+\sqrt{3}/2 \leq x \leq 2$. One can see that Monte Carlo findings are in very good agreement with our theoretical prediction Eq. (\ref{eq:Ptheta}). Dashed vertical lines corresponds to the position of the cusp singularities in Eq. (\ref{eq:cusppositions}) and it is also in excellent agreement with the Monte Carlo simulations. Note that for $x=1+\sqrt{3}/2$, cusp singularities merge in pairs, producing  only 6 singular points.

 In general, there are twelve singularities for each value of $x$, except at special points where the singularities merge in pairs. We have determined the cusp positions from the Monte Carlo data for several values of $x$, shown in Fig. \ref{baoo1}.  These are in very good agreement with the predicted values.\\ 
We parameterize the distribution function for values of $x$ outside the AOO regime  using a fitting form  with only one fitting parameter $m$, by 
\bea
P_{\text{approx}}(\theta)=\mathcal{N}\, \prod_{i=0}^5 [1+m~f(\theta -i \pi/3)],
\label{ansatz}
\eea  
where $\mathcal{N}$ is the normalization constant. Note that in the AOO regime, this expression is exact, and  the parameter  $m$ can be written in terms of the density of dimers. Outside the AOO regime, the expression is only approximate, but incorporates the known exact position of the cusp singularities, and its analytical structure is suggested by the solution of the model of interacting rods on the Bethe lattice \cite{dd2}.  The plot shown in FIG. \ref{baoo2} compares the Monte Carlo data beyond the AOO regime, for $x= 1.7 $ and $1.5$, with value of $m$ chosen to provide the  best fit.  We see  that the ansatz provides an good qualitative description for the one-point PDF. The deviations from this form occur only in the intervals of $\theta$ for which  the failure of  the AOO condition is possible. 
\begin{figure}[htp!]
    \centering
    \includegraphics[width=0.38\textwidth]{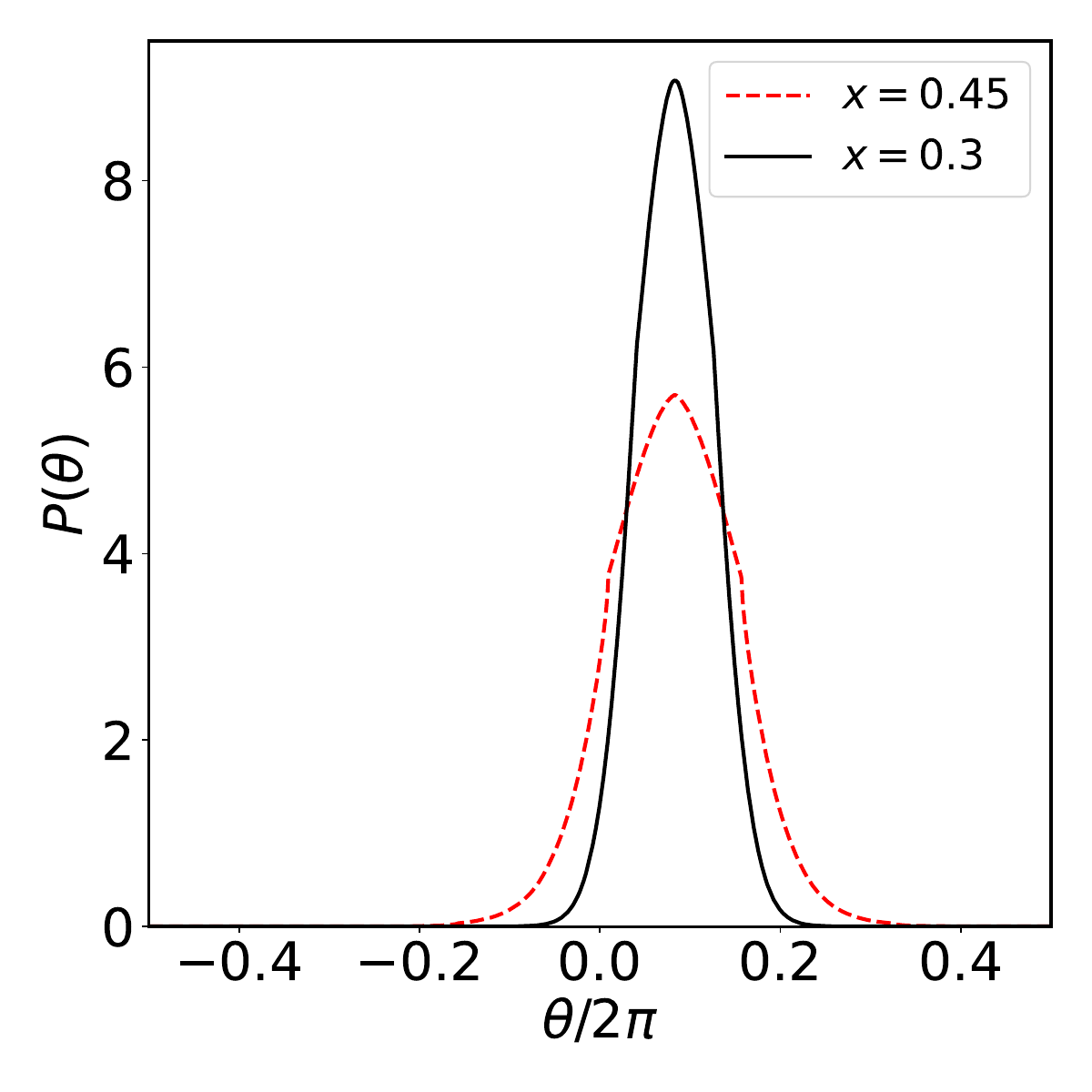}
    \caption{One-point PDF obtained using Monte Carlo simulations for $x=0.45$ and (b) $x=0.3$.}
    \label{sb1}
\end{figure}

\begin{figure}[htp]
 \includegraphics[width=0.52\textwidth]{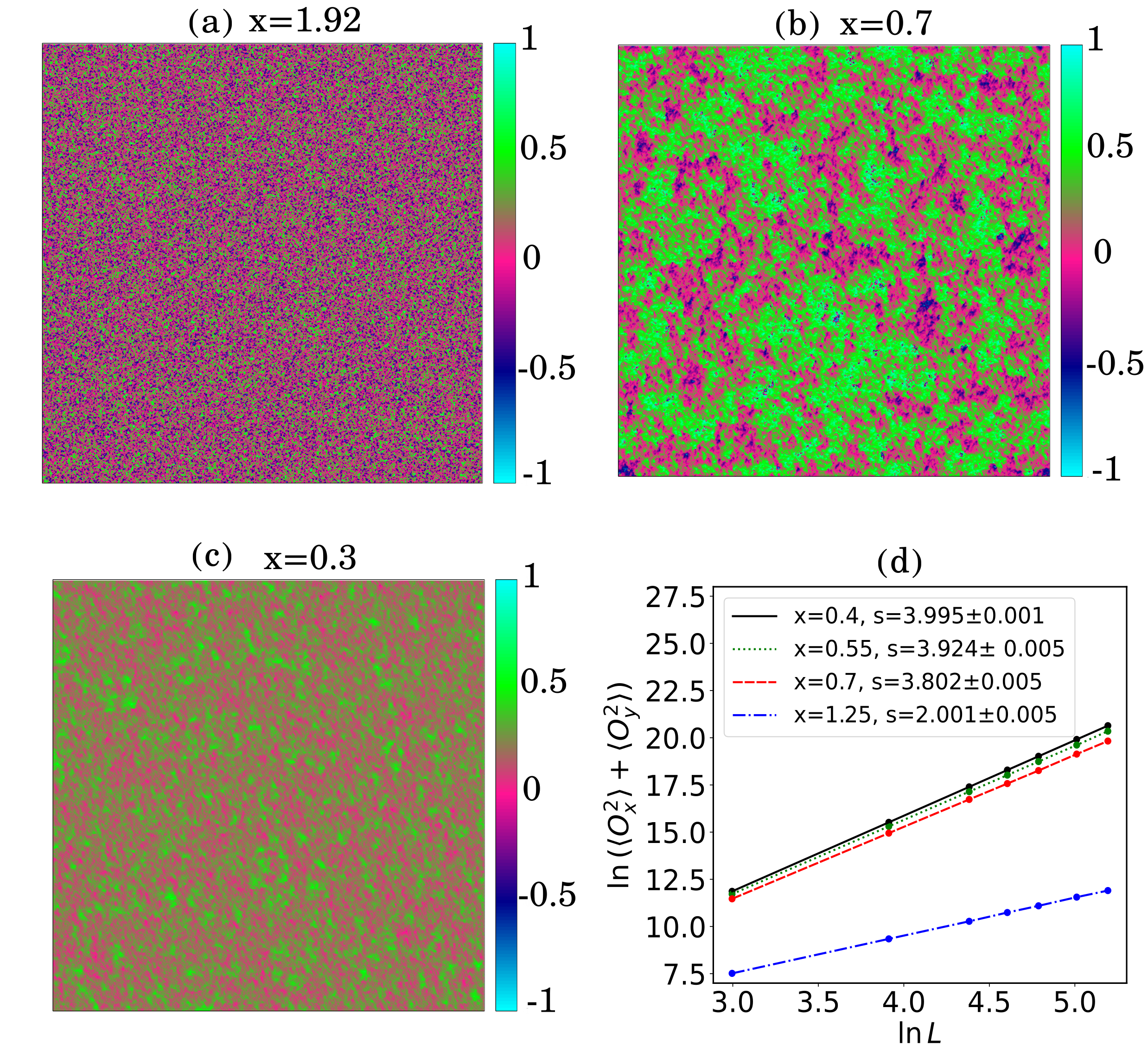}
    \caption{(a)-(c) panels show spatial heat map of orientations of discs in a 600$\times$600 triangular lattice for different $x$. Color coding is from -1 to 1 as orientations are divided by $\pi$.(a) At $x=1.92$, system is disordered as one can see all the orientations are randomly oriented giving rise to no net ordering.(b) At $x=0.7$, system is critical as cluster size of same orientations exhibits large fluctuations. (c) At $x=0.3$, system is ordered as the orientations of the discs prefers to align along a direction giving rise to net ordering.  (d) $\ln$-$\ln$ plot of mean-square of orientational order parameter vs $L$ for different $x$. $s$ represents the slope of $\ln$-$\ln$ plots for various $x$.}
    \label{phases}
\end{figure}

Our Monte Carlo simulations also revealed the presence of both the Berezinskii–Kosterlitz–Thouless (BKT) phase and the orientational-symmetry-broken phase in addition to the disordered phase. Clock-model also exhibits similar kind of behaviour where there is a intervening BKT phase between the phases with broken symmetry and the disordered phase \cite{nelson77}. A phase where the symmetry of orientational distribution function is broken under lattice rotations was observed for values of $x$ less than 0.55, as illustrated in FIG. \ref{sb1}, where the one-point probability density function is peaked at $\pi/6$ and decays rapidly to zero as we move away from it.
In the broken symmetry phase, all the singularities in $P(\theta)$ are seen only in the ensemble averaged quantities, and not in time averages of a single realization.

In the intermediate region, roughly in the range $0.55< x <0.9$, we observe BKT phase which is usually characterized by a power law decay of correlation function with the exponent that depends on $x$ and no long-range order. In FIGS. \ref{phases} (a)-(c) spatial heat maps of orientations of discs are plotted for different values of $x$ chosen from three different phases.

BKT phase can be identified by looking into the system size dependence of mean-square of orientational order parameter. On a $L\times L$ lattice, leading order dependence of mean-square of orientational order parameter on $L$ for different phases varies as,

\bea
\langle O_x^2\rangle + \langle O_y^2 \rangle \propto 
\begin{cases}
        L^2 & \text{in the disordered phase } \\
        L^4 & \text{in the ordered phase } \\
        L^{4-\eta(x)} & \text{in the BKT phase}.
    \end{cases}
\eea 
 where $O_x= \sum_{\mathbf{r}} \cos \theta(\mathbf{r}),$ $O_y=\sum_{\mathbf{r}} \sin \theta(\mathbf{r}),$ and $0 \leq \eta(x)\leq 1/4$. Angular bracket represents ensemble average. In FIG. \ref{phases}(d) mean-square of orientational order parameter vs $L$ is plotted for various values of $x$. Our simulations show that for $x > 0.9$, we get variance increasing as $L^2$, and for $x< 0.55$, it varies as $L^4$. It has an intermediate behavior, with $\eta \approx 0.08, 0.15$, and 0.20, for $x =0.55, 0.65,$ and 0.7 respectively. Nearer the BKT critical point, the relaxation becomes slow, and we are not able to get reliable estimates of the position of the critical point. We do not try to identify the range of $x$ values where the BKT phase occurs more precisely, in this paper.

\section{Concluding remarks}\label{remarks}

The model discussed in this paper is of interest for several reasons. Firstly, it is a  model simple to define,  which shows a large number of phases, and phase transitions, just by varying one parameter.  It thus provides a minimal model for describing the multitude of phases seen in plastic solids.

 Secondly, we have argued that to characterize the many phases, it is convenient to use not a single order parameter, but the  whole distribution function of orientations. The full distribution as an order parameter has been discussed in the context of spin-glasses \cite{parisi-mezard}.  Also, we have shown that this distribution function has robust geometrical singularities, whose qualitative behavior is easy to determine theoretically. To the best of our knowledge, this is first system with continuous degrees of freedom that shows non-trivial singularities in the one-point function, whose position changes when the coupling constant is varied. 

Thirdly, we could determine the angular dependence of the distribution function $P(\theta)$ for a range of values on $x$, when the AOO condition is satisfied. Outside this range, we showed that the distribution function can be expanded in a perturbation  series in a variable $y$. In general, problems where the position of singularity varies with the perturbation strength are difficult to construct.
For example, the function $P(\theta)$ is expected to have cusp singularity of the form, 
\bea
P(\theta) = A(\theta,x) + B(\theta,x) | \theta -\theta_c(x)|^{1/2},
\eea
near a cusp singularity $\theta=\theta_c(x)$, where $A(\theta,x)$ and $B(\theta, x)$ are smooth functions of $x$ (different on different sides of the cusp). A naive perturbation series in  $\delta = x - x^*$  for $P(\theta)$  about a point $x^*$ would generate spurious singularities of the type $|(\theta -\theta_c(x^*)|^{-1/2}  \delta$.  Our perturbation parameter $y$ avoids this problem, as the $x$ dependent cusp singularity structure is built in the perturbation series. 


Fourthly, the connection of this problem to the hard-disc problem is also of interest. Typical configurations generated in this model are visually not easily distinguished from the configurations of the hard-disc model at same density. In the limit $x$  tending to zero, we get the close-packed crystalline  solid. In our model, the centers of the discs cannot move freely, but each center is restricted to a circle of radius $\epsilon$. The restricted model  seems to have qualitative behavior similar to the original model. For small $x$, we do not have crystalline order, and if we construct a local coarse-grained variable  giving the average orientation of the lattice locally, this will change slowly in space, as in the hard-disc model.  In particular, for $0<x<0.5$, one can show that  there are no vortices possible. For intermediate range of $x$, we have seen that there is Kosterlitz-Thouless phase, with power-law decay of angular  correlations.

In addition to these, the system shows a series of ordering transitions. In fact, for low $x$, our Monte Carlo simulations also  show a transition to a  glassy phase with very large correlation times. We have not discussed these here. This seems to be a good direction for further studies.  

\section*{ACKNOWLEDGMENTS}
S.S. acknowledges financial support from the Department of Physics, IISER Pune. DD's work is supported by the Indian government under the  Senior Scientist Scheme of the National Academy of Sciences of India. We thank National Supercomputing Mission (NSM) for providing computing resources of `PARAM Brahma at IISER Pune’, which is implemented by C-DAC and supported by the Ministry of Electronics and Information Technology (MeitY) and Department of Science and Technology (DST), Government of India.


\renewcommand{\theequation}{A\arabic{equation}}
\setcounter{equation}{0}  
\section*{Appendix A: Singularities in $P(\theta)$}
We will show that the $f_1(\theta)$ has square-root cusp singularities at $\theta^{\pm}_{\text{cusp}}= \pm \cos^{-1}(x-1)$. If $\theta=\theta^{+}_{\text{cusp}}-\xi$ where $0 < \xi \ll 1$, then we have,
\bea
\cos \theta-x= -1+x\sqrt{2-x}\,\, \xi + O(\xi^2).
\label{pe2}
\eea
Let us define,
\bea
y=\cos^{-1}[\cos{\theta}-x].
\eea
This can also be written as $\cos y=\cos{\theta}-x=-1+x\sqrt{2-x}\,\, \xi+O(\xi^2)$. If $\xi \to 0$ then $y \to \pi $. Therefore we get,
\bea
y=\pi-\sqrt{2x}\,\,(2-x)^{1/4} \,\sqrt{\xi} + \text{higher order terms in}\,\,\, \xi \nonumber \\
\eea 
Putting this in Eq.\,(\ref{ft1}) we finally get,
\bea
f_{1}(\theta)= \sqrt{2x}\,\,(2-x)^{1/4} \sqrt{\theta^{+}_{\text{cusp}}-\theta} \,\,\,\,\,\,\, \theta \to \theta^{+}_{\text{cusp}}
\eea 
Following same lines of argument one can show that cusp singularity also exists at $\theta=\theta^{-}_{\text{cusp}}=-\cos^{-1}(x-1)$. Other $f_i(\theta)$ can also be found from $f_1(\theta)$ using the symmetries of triangular lattice, thus $f_i(\theta)=f_1\left(\theta-(i-1) \pi/3\right)$. Similarly, two cusp singularities exist for each $f_i(\theta)$ whose positions can be found using the symmetries of the triangular lattice. As a consequence, there exist twelve cusp singularities in $P(\theta)$ whose positions are given by,
\bea
\theta_{\text{cusp}}=  \frac{j\pi}{3}\, \pm \,\arccos(x-1)\,\,\, \text{where} \,\,\, j=0,1,2,....,5. \nonumber \\
\eea

\end{document}